\documentclass[a4paper]{article}
\pdfoutput=1

\usepackage{INTERSPEECH2022}

\title{WOLONet: Wave Outlooker for Efficient and High Fidelity Speech Synthesis}
\name{Yi Wang, Yi Si}
\address{Scybi}
\email{artwang007@gmail.com, yis28938@gmail.com}
\begin{document}

\maketitle
\begin{abstract}
 Recently, GAN-based neural vocoders such as Parallel WaveGAN\cite{yamamoto2020parallel}, MelGAN\cite{kumar2019melgan}, HiFiGAN\cite{kong2020hifi}, and UnivNet\cite{jang2021univnet} have become popular due to their lightweight and parallel structure, resulting in a real-time synthesized waveform with high fidelity, even on a CPU. HiFiGAN\cite{kong2020hifi} and UnivNet\cite{jang2021univnet} are two SOTA vocoders. Despite their high quality, there is still room for improvement. In this paper, motivated by the structure of Vision Outlooker from computer vision, we adopt a similar idea and propose an effective and lightweight neural vocoder called WOLONet. In this network, we develop a novel lightweight block that uses a location-variable, channel-independent, and depthwise dynamic convolutional kernel with sinusoidally activated dynamic kernel weights. To demonstrate the effectiveness and generalizability of our method, we perform an ablation study to verify our novel design and make a subjective and objective comparison with typical GAN-based vocoders. The results show that our WOLONet achieves the best generation quality while requiring fewer parameters than the two neural SOTA vocoders, i.e., HiFiGAN and UnivNet.

 
\end{abstract}

\section{Introduction}
In recent years, the development of vocoders based on deep neural networks has been rapid. Compared to conventional vocoders (\cite{morise2016world}, \cite{kawahara2006straight}), neural vocoders can significantly improve the speech synthesis quality of the current text-to-speech (TTS) system. Most early studies on neural vocoders are based on autoregressive(AR) models such as Wavenet\cite{oord2016wavenet}, WaveRNN\cite{kalchbrenner2018efficient}, SampleRNN\cite{mehri2017samplernn}, FeatherWave\cite{tian2020featherwave}, etc. In these models, the samples are generated sequentially, while the RNNs are used to model the long-term relationship in the natural waveform. Although they generate very high-quality waves, the generation speed is unfavorable due to the sequential structure, which limits their practical use in real-time TTS systems. \\
\indent To address the efficiency problem, many approaches have been proposed to accelerate the inference rate of AR models. Yu et al.\cite{yu2019durian} modify the original WaveRNN\cite{kalchbrenner2018efficient} and divide the full-band audio signal into four subbands and predict the four subband parameters simultaneously, resulting in reduced prediction time and model parameters. \cite{valin2019lpcnet} proposes another lightweight vocoder based on the WaveRNN framework. It can synthesize high-quality waveforms using linear prediction coefficients (LPC).\\
\indent Recently, non-AR models have attracted increasing attention from researchers. These models are capable of generating a waveform in a highly parallelizable manner to take full advantage of modern hardware with high inference speed. Among all these methods, the Knowledge Distill technique plays an important role (\cite{oord2018parallel},\cite{ping2018clarinet}). Specifically, the 'knowledge' of an AR teacher model is transferred to a small student model based on the inverse autoregressive fluxes combined with an additional perceptual loss. Although the resulting model can synthesize high-quality waveforms at a reasonable rate, it requires a well-trained teacher model as well as complex training strategies. The gigantic size of the model also limits its ability to compute in real-time. Another notable attempt in this area is flow-based models such as WaveGlow\cite{prenger2019waveglow}, FloWavenet\cite{kim2019flowavenet}, Melflow\cite{zeng2020melglow}, WaveFlow\cite{ping2020waveflow}, and FBWAVE\cite{wu2020fbwave}, etc. They use a single log-likelihood loss to train specially designed invertible models. The inference speed is faster than AR models and can even be used on mobile devices after additional development efforts on CPU \cite{wu2020fbwave}. However, the unstable training process and unsatisfactory synthesis quality prevent it from being used in industrial applications.\\
\indent Instead of modeling the waveform directly, WaveGrad\cite{chen2020wavegrad} uses the diffusion probability technique to model the multiscale distribution gradient field and uses Langevin dynamics to generate the waveform from random white Gaussian noise. Although the generated audio signals have high fidelity, the generation speed is greatly reduced by the multistage sampling processes.\\
\indent Some recent work has used generative adversarial networks (GAN) to train the vocoders. The generator attempts to synthesize the waveform to fool the discriminator, while the discriminator figures out the difference between the synthesized wave and the ground truth wave. When the generator reaches a Nash equalization point, it is expected to synthesize a high-quality waveform. GAN-based methods (\cite{kumar2019melgan},\cite{yamamoto2020parallel},\cite{yang2020multi},\cite{yang2020vocgan},\cite{yamamoto2020parallelv2},\cite{zeng2021lvcnet},\cite{kong2020hifi},\cite{jang2021univnet},\cite{mustafa2021stylemelgan} etc.) are promising, as some models are even capable of synthesizing waves in real-time on a single GPU or even CPU while achieving a comparable MOS that is very suitable for actual industrial use. Most of the above GAN-based vocoders differ in their network architecture and discriminator loss function. Specifically, MelGAN\cite{kumar2019melgan} first proposes the multiscale wave discriminator and feature matching loss, while ParallelWaveGAN\cite{yamamoto2020parallel} uses the WaveNet\cite{oord2016wavenet} generator structure and introduces the multi-resolution STFT loss function to accelerate the coverage, which became an important auxiliary loss in the later vocoders. MB-MelGAN\cite{yang2020multi} adopts the same idea as multi-band WaveRNN and synthesizes the sub-band signals to speed up the inference speed. Recently, HiFiGAN\cite{kong2020hifi} proposed a novel multi-period discriminator and achieved the state of the art in wave quality and real-time speed at CPU. UnivNet\cite{jang2021univnet} and Universal MelGAN\cite{jang2020universal} also propose multi-resolution spectrogram discriminators using a 2D convolution-based discriminator in the frequency domain to eliminate high-frequency artifacts, such as mental noise and reverberation in the auditory domain. StyleMelGAN\cite{mustafa2021stylemelgan} synthesizes high-quality waves by using the Adaptive Batch Normalization block conditioned by the Mel spectrogram.\\
\begin{figure}[t]
  \centering
  \includegraphics[width=\linewidth]{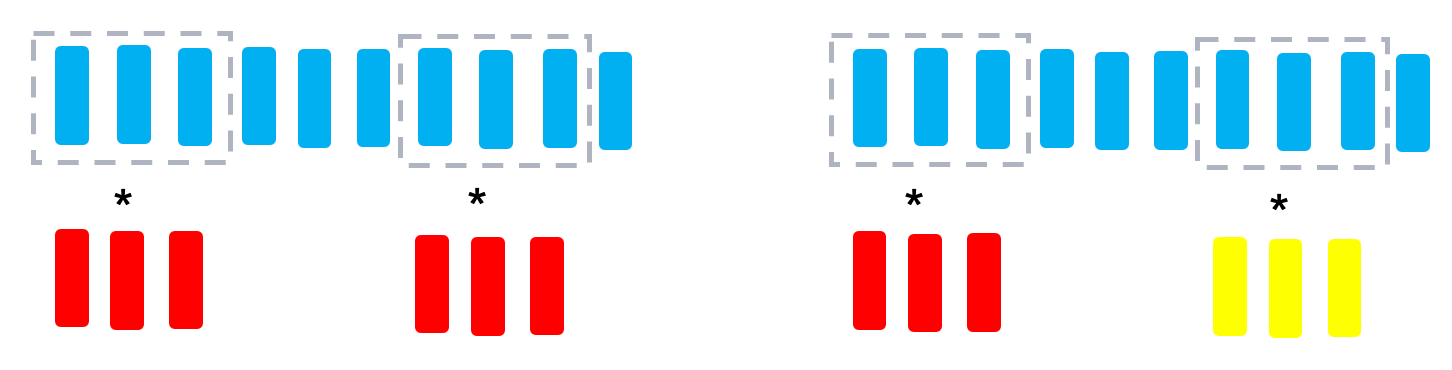}
  \caption{Most Prior works use fixed convolution kernel weights across each time step,  as shown in left, Whereas our work use various kernel weights for different time steps, i.e. location variant convolution kernel, as shown in right.}
  \label{fig:dynamic_illustrate}
\end{figure}
\indent Despite the aforementioned advances, there is still room for improvement in wave quality. We note that different wave segments may have different semantic content and phase. Using the same and fixed kernel weights for each wave segment is a suboptimal choice and the generalization ability may be limited when the trained vocoder is applied to other speakers in different environments. Therefore, location-dependent kernel weights and biases are very suitable for modeling more complex speech synthesis rules that can be used as a universal vocoder. Our main idea is to develop a lightweight and location-aware generator that can improve the quality of synthesis with fewer parameters and FLOPs. The works that are most similar to ours are LVCNet\cite{zeng2021lvcnet} and UnivNet\cite{jang2021univnet}. They both use a dynamic kernel prediction module that can generate different weights and biases for different wave segments. However, due to the limited computational budget, the hidden channel dimension is very small, i.e., 8 for LVCNet and 32 for UnivNet. The reason is due to their dynamic convolution structure, the kernel weight parameters and FLOPs are in proportion to $\mathbf{O}(k \cdot C_{in} \cdot C_{out})$, which means the kernel size and channel dimension cannot be large with the same computational budget. The shrunken channel and kernel size severely limits their modeling power. To solve this problem, we propose a novel structure WOLONet that uses location-variable, channel-agnostic, and depthwise dynamic convolution kernels. To stabilize the training process, inspired by the Fourier activation function proposed in \cite{sitzmann2020implicit} and \cite{benbarka2022seeing}, we use a sine-activated dynamic kernel and have found its superior performance over other alternative activation functions such as Softmax(\cite{yuan2021volo}) or Tanh in our ablation experiments. In our experiments, we have shown that our proposed WOLONet performs better than existing models on both subjective and objective scores with fewer parameters and FLOPs.

\section{Proposed Method}
\subsection{HiFiGAN}
HiFiGAN\cite{kong2020hifi} can generate high-fidelity waveforms in real-time on a CPU that forms the basis for our WOLONet. As a GAN-based model, HiFiGAN consists of a generator (G), a discriminator (D), and Mel spectrogram auxiliary losses and feature matching losses for faster coverage. The generator is very similar to the original MelGAN, but with several stacks of residual dilated convolutional blocks with normalized acoustic features (e.g., Mel spectrogram) as input. In HiFiGAN, a least-squares method GAN is used to minimize loss. The multiscale waveform and the proposed multi-period discriminators are used for training. The final loss is as follows:
\begin{equation}
\label{eq:hifigan_final_loss_g}
  \mathcal{L}_{G} = \sum_{k=1}^{n}\left[\mathcal{L}_{adv}(G;D_{k})+\lambda_{fm}\mathcal{L}_{FM}(G;D_{k})\right]+\lambda_{mel}\mathcal{L}_{Mel}(G)
\end{equation}

\begin{equation}
\label{eq:hifigan_final_loss_d}
  \mathcal{L}_{D} = \sum_{k=1}^{n}\mathcal{L}_{adv}(D_{k};G)
\end{equation}

\noindent where the adversal loss $\mathcal{L}_{adv}$ is the LSGAN framework:
\begin{equation}
\label{eq:hifigan_adv_loss_gd}
  \mathcal{L}_{adv}(D_{k};G) = \mathbb{E}_{(x,s)}[(D_{k}(x)-1)^{2} + (D_{k}(G(s)))^{2}]
\end{equation}
\begin{equation}
\label{eq:hifigan_adv_loss_dg}
  \mathcal{L}_{adv}(G;D_{k}) = \mathbb{E}_{s}[(1 - D_{k}(G(s)))^{2}]
\end{equation}
\noindent the feature matching loss is used to match all the feature maps of each discriminator,formulated as:
\begin{equation}
\label{eq:hifigan_fm_loss}
\mathcal{L}_{FM}(G,D_{k}) = \mathbb{E}_{(x,s)}\left[\sum_{i=1}^{T}\frac{1}{N_{i}}\|D_{k}^{i}(x) - D_{k}^{i}(G(s))\|_{1} \right] 
\end{equation}
\noindent the Mel loss is simply to match generated mel-spectrogram with the ground truth mel-spectrogram:
\begin{equation}
\label{eq:hifigan_mel_loss}
\mathcal{L}_{Mel}(G) = \mathbb{E}_{(x,s)}\left[\|\phi(x) - \phi(G(s))\|_{1} \right]
\end{equation}

\subsection{Wave Outlooker Attention Block}
\begin{figure}[t]
  \centering
  \includegraphics[width=\linewidth]{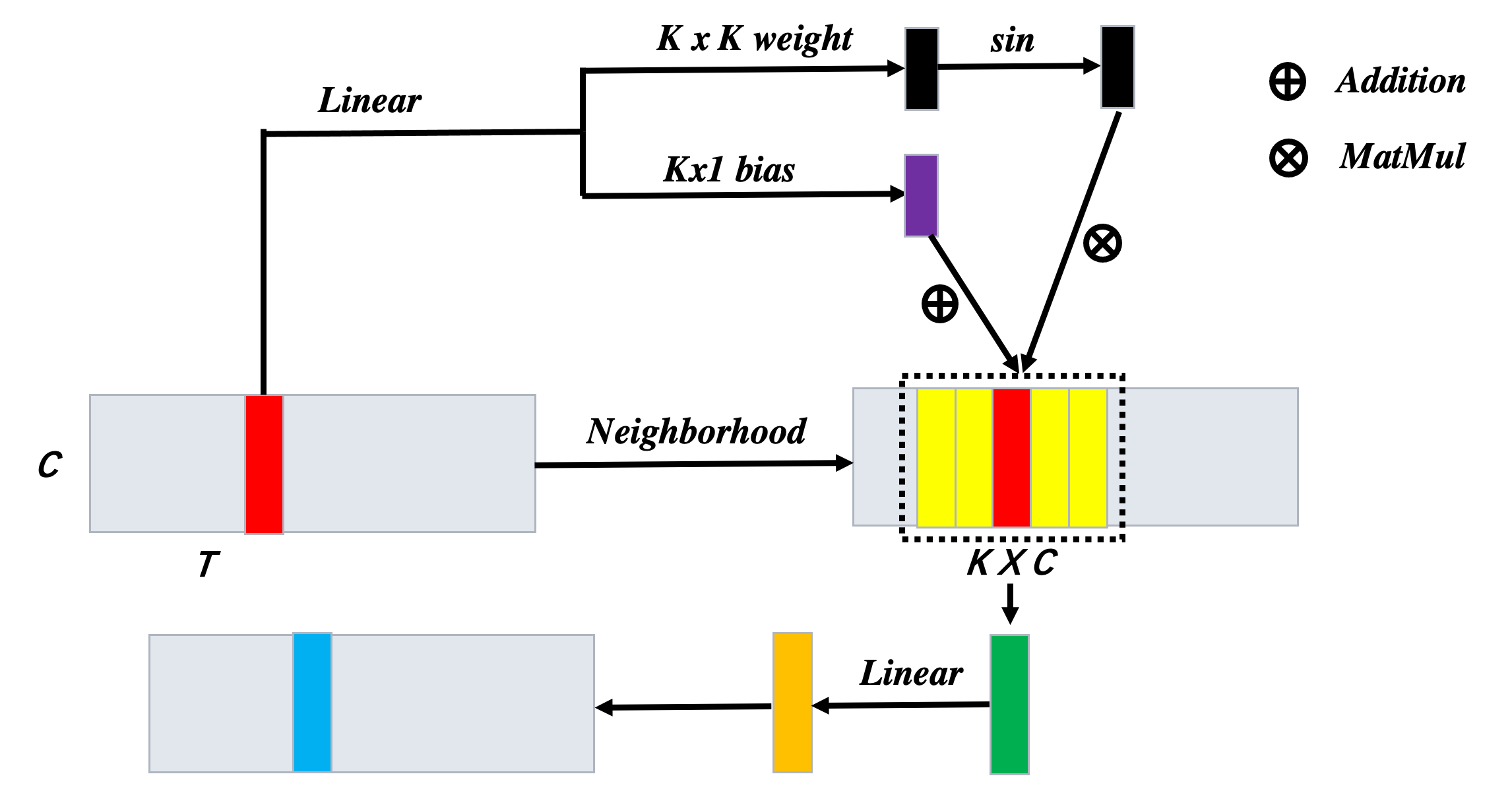}
  \caption{Structure of the proposed WOLO Attention.}
  \label{fig:wolo_block_attention}
\end{figure}

\noindent Similar to \cite{yuan2021volo}, we also call our core block outlooker attention block, but this time we will use a 1D convolution due to our 1D wave signal input. The original vision outlooker attention \cite{yuan2021volo} is used in the ImageNet classification task. As far as we know, this is the first time we use the local attention mechanism in the neural vocoder domain. In the following section, we will describe the Wave Outlooker Attention Block (WOLO block) in detail. The overall structure of the WOLO block is shown in figure \ref{fig:wolo_block_attention}. \\
\indent The Wave-Outlooker-Attention block consists of a local dilated-attention layer as a locality-variant and depthwise dynamic convolution and a post-convolution layer for mixing the channel-wise information. Note that the local dilated-attention layer uses depthwise convolutions to maintain both the efficiency and complexity of the model. Given the input wave function $X \in \mathbb{R}^{C \times T}$, we can write our WOLO block as follows: \\
\begin{equation}
\label{eq:wolo_block}
Z = X + Conv(Act(WOLOAttn(Act(X))))
\end{equation}
\noindent Here $Act(\cdot)$ is LeakyReLU used in HiFiGAN \cite{kong2020hifi}, $Conv$ is a normal 1D convolution.\\

\textbf{WOLO Attention}. The WOLO Attention is simple and efficient. The basic idea is that, first, the feature at each time step in the wave is highly representative and sufficient to generate the appropriate dynamic depth-wise weights and biases for local aggregation of neighboring wave features in the timeline, and second, local aggregation in the channel and time dimensions can efficiently encode very fine details and enlarge the resulting receptive field. \\
\indent For each time step $t$ with feature $X_{t}\in\mathbb{R}^{C}$, the Wave Outlook Attention computes its kernel weights and biases using pointwise convolution. After an activation function, the kernel weights and biases per time step are used to compute depthwise convolution at each time step $t$, and posterior pointwise convolution is used to mix the channel information similar to MobileNet (\cite{howard2017mobilenets},\cite{sandler2018mobilenetv2}). \\
\indent Mathematically, given the input $X \in \mathbb{R}^{C \times T}$ and the current time step $t$, we first calculate the kernel weight $W_{t} \in \mathbb{R}^{K \times K}$ and the bias $b_{t} \in \mathbb{R}^{K}$ for kernel size $K$ using point-wise convolution. The kernel weights are processed by the periodic activation functions (the sinusoidal activation) as proposed by \cite{sitzmann2020implicit}:

\begin{align}
\label{eq:wolo_block_kernel}
w_{t},b_{t} = UX_{t} + V \\
W_{t} = sin(w_{t})
\end{align}
$U$ and $V$ are pointwise convolution weight and bias, which are shared in all time steps. We then consider the $K$ neighborhood of $X_{t}$, denoted as $Y_{t} \in \mathbb{{R}^{K \times C}}$:
\begin{equation}
\label{eq:wolo_block_nh}
Y_{t} = \left\{X_{t + p - \left \lfloor \frac{K}{2} \right \rfloor}\right\},0 \le p < K
\end{equation}
\noindent Note here the neighborhood definition will change according to the specified dilation factor, we ignore it in the equation for brevity.\\
\indent Then we can compute the transformed convolution values $\hat{Y}_{t} \in \mathbb{R}^{K \times C}$ using the matrix multiply:
\begin{equation}
\label{eq:wolo_block_matmul}
\hat{Y}_{t} = MatMul(W_{t},Y_{t})+b_{t}
\end{equation}
\indent The \textbf{dense aggregation} operation is used to obtain the attention feature $\bar{Y}_{t} \in \mathbb{R}^{C}$:
\begin{equation}
\label{eq:wolo_block_aggregate}
\bar{Y}_{t} = \sum_{0 \le p < K}\hat{Y}_{t + p - \left \lfloor \frac{K}{2} \right \rfloor}
\end{equation}
\noindent Actually, the dense aggregation is implemented by the \textit{fold} function in the PyTorch. \\
\textbf{Discussion.} \noindent From the block construction approach, we can calculate the multiplicity addicts (M-Adds) for the WOLO Attention Block, approximately $2(K^{2}+K)C + C^{2} + C$, for a typical structure such as the residual block in HiFiGAN, M-Adds is approximately $KC^{2} + C$. Thus, when $C=512$,$K=5$, the M-Adds reduction ratio is about $\frac{1310720}{293376}=4.468$, i.e., about a fourfold reduction, which theoretically shows that our proposed WOLO block is lightweight and efficient. Another advantage of the WOLO block over alternative structures such as LVCNet\cite{zeng2021lvcnet} or UnivNet\cite{jang2021univnet} is that the dimensions of the location variant weights and biases are channel-independent, so we can easily maintain the channel size with a large number and preserve model complexity.
\subsection{Wave Outlooker Network}
\begin{figure}[t]
  \centering
  \includegraphics[width=\linewidth]{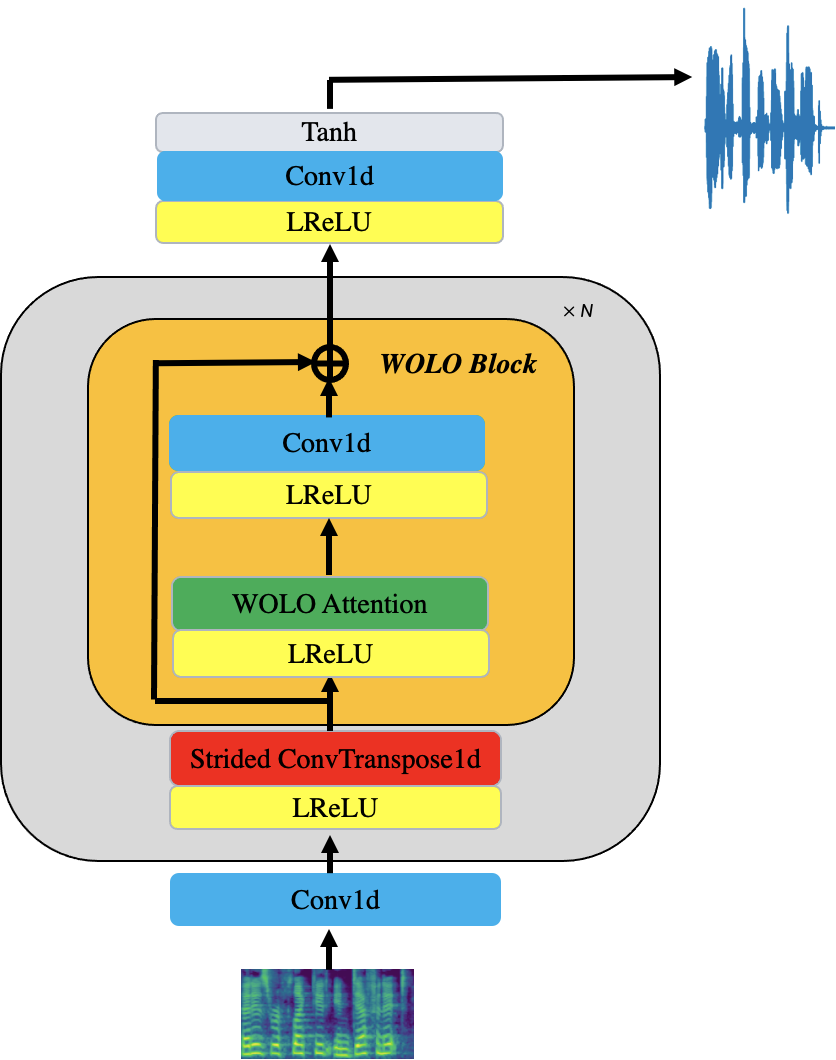}
  \caption{Structure of the proposed WOLONet.}
  \label{fig:wolo_net}
\end{figure}
Based on the Wave Outlooker Attention Block described in the previous section, we now describe our proposed WOLONet. As shown in Figure \ref{fig:wolo_net}, our WOLONet is similar to the HiFiGAN structure, except that the remaining dilated convolutional blocks are replaced by our WOLO block. The discriminator structure and training loss are the same as HiFiGAN. We have found that this is sufficient for the high quality of WOLONet.
\begin{table*}[!htbp]
\centering%
\begin{tabular}{c|cccc|}
  \toprule
  Activation & MCD($\downarrow$) & PESQ(wb) & PESQ(nb) & MOS \\
  \midrule
  Softmax & 2.45 & 3.34 & 3.67 & 3.78$\pm$0.17\\
  Tanh & 2.35 & 3.55 & 3.82 & 3.96$\pm$0.12\\
  Sine & \textbf{2.32} & \textbf{3.62} & \textbf{3.87} & \textbf{4.02}$\pm$\textbf{0.11}\\
  Recordings & 0.0 & 4.5 & 4.5 & 4.22$\pm$0.07 \\
  \bottomrule
\end{tabular}
\captionsetup{justification=centering,margin=2cm}
\caption{Ablation study on the choice of activation function in WOLO block.}
\label{tab:results_ablation}
\end{table*}

\begin{table*}[!htbp]
\centering%
\begin{tabular}{c|ccc|c|c|c}
  \toprule
  Method & MCD(dB) & PESQ(wb) & PESQ(nb) & Param(M) & RTF & MOS \\
  \midrule
  HiFiGAN V1\cite{kong2020hifi} & \textbf{2.286} & 3.46 & 3.74 & 13.29 & 0.014 & 3.87$\pm$0.15\\
  FB-MelGAN\cite{yang2020multi} & 2.522 & 3.03 & 3.47 & 4.07 & \textbf{0.009} & 3.51$\pm$0.16 \\
  Paralle-WaveGAN\cite{yamamoto2020parallel} & 2.94 & 2.97 & 3.39 & \textbf{1.28} & 0.048 & 3.41 $\pm$0.15 \\
  UnivNet32c\cite{jang2021univnet} & 2.55 & 3.47 & 3.78 & 14.16 & 0.024 & 3.91 $\pm$0.12\\
  WOLONet & 2.32 & \textbf{3.62} & \textbf{3.87} & 9.09 & 0.027 & \textbf{4.02} $\pm$\textbf{0.11} \\
  Recordings & 0.0 & 4.5 & 4.5 & N/A & N/A & 4.22$\pm$0.07 \\
  \bottomrule
\end{tabular}
\centering
\captionsetup{justification=centering,margin=2cm}
\caption{Comparison against alternative methods. The results of subjective and objective scores are listed.}
\label{tab:results}
\end{table*}

\section{Experiments}
\subsection{Experiments Setup}
For experiments, we use an open-source studio-quality LJSpeech dataset \footnote{available at http://data.keithito.com/data/speech/LJSpeech-1.1.tar.bz2}, which contains 13100 audio samples from an English female speaker. The total length of the audio samples are 24 hours. All the recordings have been resampled to 22050 Hz sampling rate with the PCM-16-bit format. Follow the recipe of Parallel WaveGAN\footnote{https://github.com/kan-bayashi/ParallelWaveGAN}, We randomly selected 250 utterances from the corpus as a validation set and another 250 utterances as our test set. All remaining samples are used for training. We extracted 80 dim log-Mel-spectrograms by 1024-point Fourier transform with 256 hop lengths. with the minimum and maximum frequency as 80 and 7600. We conducted our experiment on a server with a single NVIDIA Tesla V100 GPU for training and evaluation.
\subsection{Implementation details}
For the HiFiGAN structure, we strictly follow the original implementation, and use HiFiGAN v1 as our baseline structure. Concretely, The generator consists of four transposed convolutional blocks where each block has stride 8, 8, 2,2 respectively. Each transposed block contains 3 residual blocks with dilated convolutions, and their dilation factors are 1, 3, 5. For the discriminator, we use the exact implementation from Parallel  WaveGAN Code\footnote{https://github.com/kan-bayashi/ParallelWaveGAN}.\\
\indent Adam\cite{kingma2014adam} is chosen as the optimizer for HiFiGAN. As for the learning rate configuration, the learning rate of a generator was initialized to 2e-4, reducing by half every 200K iterations. The discriminator was also set to 2e-4 and halved at every 200K iterations. The total training iteration is set to 800K, and batch size is set to 16. For MelGAN and ParallelWaveGAN, we use the exactly same recipe from \footnote{https://github.com/kan-bayashi/ParallelWaveGAN}. For UnivNet\cite{jang2021univnet}, we strictly follow the training recipe in \footnote{https://github.com/mindslab-ai/univnet}, i.e., AdamW as the optimizer and the upsampling factor is 8,4,4. Other training configurations are the same as HiFiGAN for a fair comparison. \\
\indent We have used two objective metrics and one subjective metric to evaluate our methods. To measure the accuracy of the synthesized waveform, we measured the Mel Cepstral Distortion (MCD) proposed in \cite{skerry2018towards}. For the evaluation of synthesized waveform quality, we measured narrow and wide band PESQ\cite{rix2001perceptual} scores (Note the audios are downsampled to 16kHZ before evaluating PESQ score), and use 5-point(1 point is the worst and 5 point is the best) Mean Opinion Score (MOS) to evaluate subjective synthesis quality. To obtain the subjective MOS, we ask 20 subjects to evaluate the synthesized and recording audio samples. the assessed scores
were collected to compute the mean and 95\% confidence intervals for each assessment. 
\subsection{Evaluation}

\textbf{Ablation Study.} To validate our choice of periodic activation function, we compared the performance of three alternative activation functions, softmax\cite{yuan2021volo}, Tanh, and the sinusoidal function used in WOLONet. The results from the table \ref{tab:results_ablation} clearly show the superior performance of the sinusoidal function. It is known that the softmax activation function is commonly used for the self-attention mechanism, while in speech synthesis, the wave signals seem to prefer the periodic activation function, which matches the periodic property of the signal itself. Softmax activation largely limits the expressive power of WOLONet and Tanh is a sub-optimal choice for our WOLONet. Therefore, we propose that the periodic activation function could be very useful in other speech synthesis tasks, such as text-to-speech, which is our future work.\\

\textbf{Comparison Against Alternative Methods.} WeWe conducted experiments to evaluate the effect of our proposed methods on the LJSpeech dataset. Our experiment is called Mel2Wav,i.e. synthesizing the wave using the ground truth Mel-spectrogram. For this experiment, we use the 250 test utterances as the input to the alternative vocoders as well as our model and measure the average MCD, PESQ, and the MOS score for each method and the recording audios. Table 2 demonstrates the results. As proved in \cite{jang2021univnet}, UnivNet-32 achieves a better score than HiFiGAN vocoders, but the improvement is small in our limited bandwidth dataset(80HZ-7600HZ). However, the WOLONet achieves the optimal 
PESQ score in both narrow and wide bands with a margin of 0.09 over the current UnivNet-32, and the parameter number is still less than HiFiGAN and UnivNet-32, which demonstrates that WOLONet is a lightweight, efficient, and high fidelity neural vocoder. Other vocoders such as MelGAN and Parallel WaveGAN, despite their low computational demand and fewer parameters, have inferior wave quality compared with HiFiGAN, UnivNet, and our WOLONet. Note that the RTF(realtime factor) value for the WOLONet is larger than UnivNet and HiFiGAN, the main reason lies in the slow implementation of \textit{fold} operation in PyTorch, However, when written in an optimized CUDA kernel, the inference speed will be faster than HiFiGAN and UnivNet, since we have theoretically analyzed the computational complexity of WOLONet and HiFiGAN in the previous section. Above all, when compared with alternative GAN-based vocoders, WOLONet can generate higher quality waves while keeping the parameter number small for the actual deployment.
\section{Conclusions}
In this work, we proposed WOLONet, a lightweight and efficient neural vocoder whose core is a location-variant, channel-independent, and depthwise dynamic convolutional kernel with a sinusoidally activated dynamic kernel weight. Our experimental results have shown that WOLONet can achieve higher subjective and objective scores when compared against alternative baseline methods with fewer parameters and FLOPs. In the future, we will train our WOLONet with more different training datasets and develop a universal neural vocoder that can be used for various speakers and heterogenous environments.

\bibliographystyle{IEEEtran}

\bibliography{WoloNet.bbl}


\end{document}